\documentclass[aps,twocolumn,pra,showpacs,floatfix]{revtex4}
\usepackage{epsfig}
\usepackage{graphicx}
\usepackage{dcolumn}

\begin{document}

\title{Exponential increase of energy level density in atoms: Th and Th~II}
\author{V. A. Dzuba and V. V. Flambaum}
\affiliation{School of Physics, University of New South Wales, Sydney 2052, 
Australia}

\date{\today}

\begin{abstract}
We present analytical estimates and numerical calculations showing that
the energy level density in open-shell atoms increases exponentially
 with increase of excitation energy. As an example,
we use the relativistic Hartree-Fock and configuration interaction methods
to calculate the density of states of Th and Th~II. The result is
used to estimate the effect of electrons on the nuclear
transition which is considered for the use in a nuclear clock. 

\end{abstract}


\maketitle

\section{Introduction}

The exponential increase of density of states is well known in nuclear physics
(see, e.g.~\cite{Bohr}). Similar problem in atomic physics has not
been considered so far. It is well-known that the density of the Rydberg levels
(with one excited electron) tends to infinity at ionization limit. 
 However, there are also compound states with several excited electrons.
 As we will see below, the density 
of such states increases exponentially with the excitation energy.

The concept of the density of states $\rho (E)$
can be especially  useful for the cases of the dense and complicated
spectrum. This takes place for practically any many-electron atom at
sufficiently high excitation energies, in particular at energies close
or above the ionization limit. In the latter case we speak about
density of Feshbach resonances corresponding to the quasi-stationary compound states
with several excited electrons. For atoms with open $d$ or $f$ shells the
spectrum is complicated even at low energies. Experimental data for
highly excited states are often absent. Accurate calculations
using specific atomic states in the area of dense spectrum is very
difficult if not impossible. Even identification of such states is
often problematic. On the other hand, statistical approaches, which
use functions obtained by averaging over large number of states, can
be very useful. An interesting case was considered in
Ref.~\cite{Ce} where properties of excited states of cerium atom were
studied using statistical analysis. It has been demonstrated that the
Ce atom is an example of quantum chaotic system in which highly
excited states are similar to compound states of heavy nuclei.  
Density of states $\rho$ is used in the criteria for the chaos to take place:
$H_{ij} > D_{ij} \equiv 1/\rho_{ij}$. Here $H_{ij}$ is the off-diagonal matrix
element of the effective Hamiltonian between many-electron states $i$
and $j$, and $D_{ij}$ is the energy interval between the states
which can be mixed by  $H_{ij}$.
Statistical theory based on the properties
of chaotic eigenstates \cite{Ce,Stat} allows one to calculate average
orbital occupation numbers, matrix elements and amplitudes between chaotic
 states, enhancement of weak interactions and increase of entropy.
 Another important example is the effect of
atomic electrons on nuclear transitions, which will be considered below.
 The concept of the density of states is also needed to estimate
a number of states missed in experimental spectra.

In this paper we present simple analytical estimates explaining exponential
increase of density of states containing  several excited electrons.
Then we perform numerical calculations of  density of compound states
using thorium as an example due to its importance for the
construction of nuclear clock.  

The proposed clock~\cite{thclock} utilizes the use of
the nuclear transition between a
metastable isomeric state and the ground state of $^{229}$Th. The
frequency of this transition is unusually low, about 7 eV,
 and can be accessed by
lasers. Additional interest is due to the fact
that this transition must be very sensitive to the time-variation of
the fine structure constant~\cite{Flambaum}. The effect of electrons
on the nuclear transition in Th II and Th IV ions, also known as
 {\it electronic bridge}
process, was considered in Refs.~\cite{Porsev1,Porsev2}
(see also \cite{bridge}). In this process the
nuclear transition
 is accompanied
by an electron transition from the ground state to an exited state followed by
emission of the photon. The electronic bridge process may have significant
 effect on the performance of the nuclear clock in the case of resonance
 between nuclear and atomic transitions. However, neither nuclear frequency nor
electron spectrum are known sufficiently well to make reliable
estimations. The measured values for the frequency of the nuclear
transition varies form $3.5 \pm 1.0$ eV~\cite{th1} to $7.6 \pm
0.5$~\cite{th2}. Electron spectrum of Th~II is also not known at energies
around 7.6 eV. In this situation instead of calculating the value of
the electronic bridge process one can consider estimation of the
probability for the result to be within some range of values. This
can be done if density of electron states as a function of the energy
is known.  Knowledge of the electron spectrum density in  Th~II
is also needed for the accurate measurements of the nuclear transition
frequency. 
E. Peik and collaborators~\cite{Peik} plan to use laser excitation
of a  Th~II electron state which may transfer part of the excitation
energy to induce the nuclear transition.


\section{Simple analytical estimate of the number of compound states}

Let us consider an open ground electron  shell containing $n$ electrons
and $g_0$  single-electron states; usually $g_0=\sum_l 2(2 l+1)$ where
$l$ is the orbital angular momentum of a single-electron state within
this shell. A total number of many-body states within this shell is equal
to the binomial coefficient,
\begin{equation}\label{N0}
N_0= C^n_{g_0}=\frac{g_0!}{n!(g_0-n)!}\approx
\frac{ \exp {[n \ln{(g_0/n)}+1]}}{\sqrt{(2 \pi n)}}  .
\end{equation} 
In the last expression we  assumed $g_0 \gg n $ and used the Stirling formula
\begin{equation}\label{Stirling}
 n!\approx \sqrt{2 \pi n} \, n^n e^{-n} 
\end{equation} 
which gives an excellent accuracy even for $n=1$ (the correction is $-1/(12n)$).
Assume for simplicity that all these many-body states have approximately
 the same energy.  Let us now consider 
 next electron shell which is separated from the ground state shell by a
single-particle energy interval $\omega$ and has $g_1$ 
 single-electron states . If a total excitation energy
of a many-electron state is equal to $E$, we may transfer $k=E/\omega$
 electrons
from the ground shell to the next shell. Now we have a total number of
 many-electron states    
\begin{equation}\label{NE}
N(E)= C^{g_0}_n+ C^n_k C^{g_0}_{n-k} C^{g_1}_k \,.
\end{equation} 
Here the first factor $ C^n_k$ gives the number of ways to select
$k$ excited electrons from the available $n$ electrons,
the second factor gives the number of many-body states in
 the ground shell and the last factor gives the number of many-body states
in the next shell.
For $n \gg k$ 
\begin{eqnarray}\label{Nexp}
\nonumber
N(E) &\approx& N_0\left(1+\frac{ \exp {[k \ln{(n^2g_1/k^2g_0)}+2k]} }{2
    \pi k}\right) \\
&=&N_0+ \frac{a}{E}\exp{(bE)} \,,
\label{eq:exp}
\end{eqnarray} 
where
\begin{equation}\label{a}
 a\approx \omega \frac{ \exp {[n \ln{(g_0/n)+1]} }}{2 \pi \sqrt{(2 \pi n)}}
\end{equation}
\begin{equation}\label{b}
 b\approx  \frac{ \ln{[(n^2 g_1)/( k^2 g_0)]}+2}{\omega}
\end{equation}
This rough estimate shows that the number of compound states increases
 exponentially with the number of electrons in the open shell $n$ and with the
number of excited electrons $k=E/\omega$.
Inclusion of the electron excitations to higher shells and the Rydberg spectrum
of the single-electron excitations will only make the number of states larger.

The derivation of formula (\ref{eq:exp}) assumes that the number of
electrons $n$ in the open shell is large. We will see below that the
exponential increase of the density of states takes place even for $n$
as small as three.

\section{Numerical calculations}

\begin{table} 
\caption{Number of states of even parity and total momentum $J$=1.5 in
  10000~cm$^{-1}$ energy intervals in Th II.}
\label{t:nos}
\begin{ruledtabular}
\begin{tabular}{rcrrrrr}
\multicolumn{3}{c}{Interval} & 
\multicolumn{1}{c}{Exp.\footnotemark[1]} & 
\multicolumn{1}{c}{Full CI\footnotemark[2]} & 
\multicolumn{1}{c}{$H_{ii}$\footnotemark[3]} & 
\multicolumn{1}{c}{$H_{ii}$\footnotemark[4]} \\
\hline
    0 & - &  10000 &   5 &   5  &   1  &   3 \\
10001 & - &  20000 &   3 &   3  &   5  &   5 \\
20001 & - &  30000 &   4 &   5  &   4  &   7 \\
30001 & - &  40000 &  10 &  11  &   7  &  15 \\
40001 & - &  50000 &  18 &  10  &  19  &   8 \\
50001 & - &  60000 &   8 &  18  &   5  &  26 \\
60001 & - &  70000 &     &  50  &  44  &  90 \\
70001 & - &  80000 &     &  93  &  94  & 160 \\
80001 & - &  90000 &     & 161  & 173  & 300 \\
90001 & - & 100000 &     & 128  & 142  & 168 \\
\end{tabular}
\end{ruledtabular}
\footnotetext[1]{Experiment~\cite{th-el}}
\footnotetext[2]{Short basis}
\footnotetext[3]{No diagonalization, short basis}
\footnotetext[4]{No diagonalization, long basis}
\end{table}

Density of atomic states $\rho$ can be defined via the relation
\begin{equation}
  N(E) = \int_0^{E} \rho(\epsilon) d\epsilon,
\label{eq:rho}
\end{equation}
where $N(E)$ is the number of atomic states in the interval from
zero energy, which corresponds to the ground state, to the energy
$E$. Total density $\rho$ is the sum of the partial densities
$\rho_p$ which correspond to the states of definite total momentum $J$
and parity:
\begin{equation}
  \rho(\epsilon) = \sum_p \rho_p(\epsilon).
\label{eq:srho}
\end{equation}
We use relativistic Hartree-Fock (RHF) and configuration interaction
(CI) methods to calculate the number of states for given $J$ and
parity. The partial density is calculated by the numerical differentiation
\begin{equation}
  \rho_p(\epsilon) =
  \frac{N_p(\epsilon+\delta)-N_p(\epsilon-\delta)}{2\delta}. 
\label{eq:nrho}
\end{equation}
The value of $\delta$ is chosen to have a smooth function for
$\rho_p$. This is achieved when the number of states in the numerator
is large. The density function defined this way represents inversed
energy interval $D_p$ averaged over large number of neighboring
states, $\rho_p=1/D_p$.

We perform the configuration interaction (CI) calculations using the $V^{N-1}$ basis
(see e.g.~\cite{Kelly}) calculated in the frozen  RHF
field of  the [Ra]$6d^2$ configuration for Th~II and the
[Ra]$6d^27s$ configuration for Th~I.
To test the accuracy of the calculation we use two different basis
sets for Th~II: a short one and a long one.
The purpose of using short basis is to show that the
non-diagonal matrix elements of the CI matrix can be neglected in the
evaluation of the density of states. This  allows us to use much
larger basis and, consequently, move to higher energies.

 In the short basis we
include six lowest RHF states above the core in each of the partial
waves up to $l_{max}=4$. Two more states in each partial wave are
added in the long basis. Apart from that, we allow only single and
double excitations from a reference configuration to construct
three-electron states for the CI calculations with the use of the
short basis. Three-electron excitations are allowed when long basis is
used. 

For the short basis we perform both the full-scale CI calculations which
include matrix diagonalization, and the simplified
calculations in which off-diagonal matrix elements of the CI matrix
are neglected. This corresponds to an approximation
 $ E_M \approx \langle M|\hat H^{CI}| M \rangle$,
where $|M \rangle$ is a three-electron CI state of a definite total
momentum $J$ and parity, $\hat H^{CI}$ is the CI Hamiltonian. For the
long basis we neglect the off-diagonal matrix elements.
The use of this approximation allows us to deal with large number of
states ($\sim 10^5$). On the other hand, neglecting the non-diagonal CI
matrix elements has little effect on the density of states.

Table~\ref{t:nos} shows experimental and calculated number of states
in the 10000~cm$^{-1}$ energy intervals of Th~II. We consider even
states of total momentum $J$=1.5 as an example. These states are
important for the electronic bridge process~\cite{Porsev2}. 
Experimental data ends at about 56000~cm$^{-1}$. Next two columns
correspond to exactly the same CI matrix obtained with the use of the
short basis set. It has been diagonalised in the full CI calculations
while only diagonal matrix elements are considered in the next
column. One can see that neglecting the off-diagonal CI matrix elements
does not lead to dramatic changes in the density of states at high energies
 where the density function is well
defined (see eq.~(\ref{eq:nrho}) and discussion below it).

The last column of Table~\ref{t:nos} shows density of states obtained
with the use of long basis without diagonalization. The use of long
basis adds more states at large energy and has little effect
on the number of states at low energy. Therefore, we use only long
basis in the calculations for Th~I.

\section{Results}

\begin{figure}
\centering
\epsfig{figure=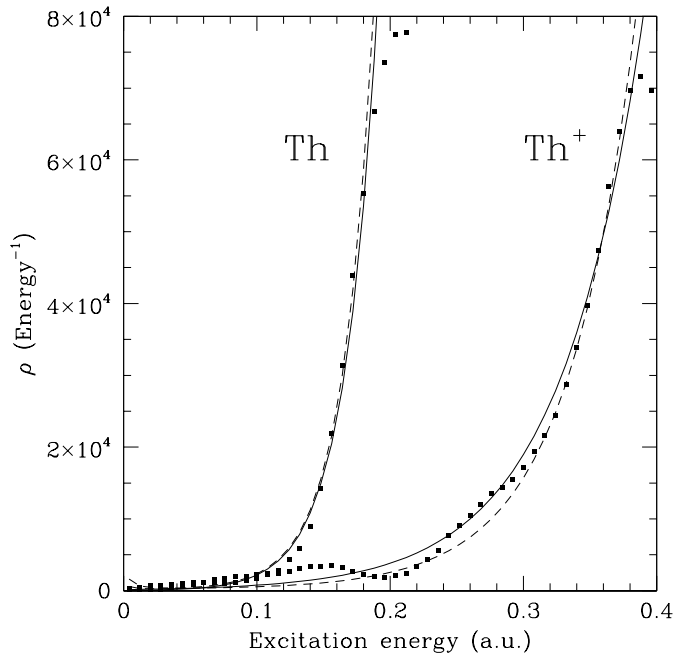,scale=0.45}
\caption{Total density of states of Th~I and Th~II as functions of excitation
  energy. Points show the result of the CI calculations, solid line -
  eq.  (\ref{eq:e1}), dashed line - eq.  (\ref{eq:e2}). The density in Th I is higher
since the number of electrons in the open shell is larger, $n=4$,  and the interval
between the single-particle levels $\omega$ is smaller - see eq. (\ref{Nexp}).}  
\label{f:tro}
\end{figure}

The results of the calculations for the total density of states are
shown on Fig.~\ref{f:tro}. It is also shown that the
calculated data is fitted very well by simple exponential functions. We use two
fitting formulae
\begin{equation}
  \rho(\epsilon) = \frac{A}{\epsilon_0} \exp(\epsilon/\epsilon_0)
\label{eq:e1}
\end{equation}
and
\begin{equation}
  \rho(\epsilon) = \frac{A}{\epsilon} \exp(\epsilon/\epsilon_0),
\label{eq:e2}
\end{equation}
where $A$ and $\epsilon_0$ are fitting parameters. Formula
(\ref{eq:e2}) is more consistent with the analytical formula
(\ref{eq:exp}) while formula (\ref{eq:e1}) is simpler and behaves
better at $\epsilon \rightarrow 0$.
The values of the fitting parameters $A$ and $\epsilon_0$ for the total
density of states of Th~II and Th~I are presented in Table~\ref{t:tot}.
It turns out that $\epsilon_0$ can be kept the same for all partial
densities. The values of parameter $A_p$ for partial 
densities are presented in Table~\ref{t:fit}.

\begin{table}
\caption{Fitting parameters $A$ and $\epsilon_0$ (a.u.) for total
  densities of states of Th~II and Th~I.}
\label{t:tot}
\begin{ruledtabular}
\begin{tabular}{l cccc}
 & \multicolumn{2}{c}{Formula~(\ref{eq:e1})} 
 & \multicolumn{2}{c}{Formula~(\ref{eq:e2})} \\
 & $A$ & $\epsilon_0$ & $A$ & $\epsilon_0$ \\
\hline
Th II & 9.8 & 0.0625 & 6 & 0.045 \\
Th I  & 1.06 & 0.025 & 2 & 0.021 \\
\end{tabular}
\end{ruledtabular}
\end{table}

\begin{table*} 
\caption{Fitting parameters $A$ (formula (\ref{eq:e1})) for partial
  densities of states of Th~II and Th~I.}
\label{t:fit}
\begin{ruledtabular}
\begin{tabular}{cc ccccccc cccccccc c}
 & & \multicolumn{7}{c}{Even states} & 
   \multicolumn{8}{c}{Odd states} & \multicolumn{1}{c}{Total} \\
\hline
 Th~II & $J$ & 0.5 & 1.5 & 2.5 & 3.5 & 4.5 & 5.5 & 
             & 0.5 & 1.5 & 2.5 & 3.5 & 4.5 & 5.5 & & & \\
      & $A$  & 0.6 & 0.9 & 1.2 & 1.2 & 0.9 & 0.7 & & 
              0.4 & 0.8 & 0.9 & 0.9 & 0.8 & 0.5 & & & 9.8 \\
\hline
Th~I & $J$ & 0 & 1 & 2 & 3 & 4 & 5 & 6 & 0 & 1 & 2 & 3 & 4 & 5 & 6 & 7
&  \\
      & $A$ & 0.03 & 0.07 & 0.08 & 0.07 & 0.07 & 0.05 & 0.02 &
              0.03 & 0.09 & 0.13 & 0.14 & 0.12 & 0.09 & 0.05 & 0.02 &
              1.06 \\
\end{tabular}
\end{ruledtabular}
\end{table*}

The density of states must go to infinity when energy is approaching
ionization potential. This is due to the infinite number of Rydberg states
which all have about the same energy, close to the ionization
potential. In contrast, 
the number of calculated states goes down at large energies. 
This is the
cut-off effect of the single-electron basis. Adding more highly
excited orbitals into the basis increases the number of many-electron
states at high energy (see Table~\ref{t:nos}). 
Ionization potential of Th~I is
50867~cm$^{-1}$ = 0.23~a.u.~\cite{Thiop} while the one for Th~II is
about 96000~cm$^{-1}$ = 0.44~a.u.~\cite{ThIIiop}. As can be seen from
Fig.~\ref{f:tro} the fitting works very well for the energies very
close to the ionization potentials of Th~II and Th~I.
At higher energies more basis states must be included.

Now we can use the density function to estimate the probability of the
electronic bridge process in Th~II. According to Ref.~\cite{Porsev2} the
ratio $\beta$ of the electron bridge width  to the width of the direct
nuclear transition (the electron enhancement factor) is
\begin{equation}
  \beta \approx \left(\frac{\omega}{\omega_N}\right)^3
\frac{R_n}{3(2J_n+1)(2J_i+1)(\omega_{in}+\omega_N)^2},
\label{eq:g1}
\end{equation} 
where $n$ is the resonance electron state, $\omega$ is the photon frequency
, $\omega_N$ is the frequency of the nuclear
transition, $J_n$ is the total angular momentum of the resonance
state, $J_i$ is the total angular momentum of the ground state
($J_n=J_i$=1.5 in our case), $R_n$ is the combination of the electron
matrix elements for electric dipole and hyperfine transitions between
atomic ground state, resonance state $n$ and some odd-parity low lying
final atomic state, $\omega_{in}$ is the frequency of the electron
transition from the ground to the resonance state, $\omega_N$ is the
frequency of the nuclear transition. Resonance corresponds to the
condition $\omega_{in} \approx - \omega_N$. As it was discussed in the
introduction the exact value for the nuclear frequency is not known
and the experimental values range from 3 eV to 
8~eV. Estimation  of the effect for 3.5 and 5.5 eV were performed in
Ref.~\cite{Porsev2}. The effect for higher $\omega_N$ was not presented since
the electron spectra are not known for such energies.
 Here we perform the estimation for the latest measured value $\omega_N$ =
7.6~eV = 0.28~a.u. Using formula (\ref{eq:e1}) and the data from
Table~\ref{t:fit} we find that the average energy interval between
the relevant electron states  is
$D_p=1/\rho_p = 6\times 10^{-4}$~a.u  at this energy. The energy denominator in
(\ref{eq:g1}) is at least two times smaller. Therefore,
the electron enhancement factor is
\begin{equation}
  \beta \sim 2 \times 10^5 R_n \approx 4 \times 10^3.
\label{eq:g1e}
\end{equation}
Here we use the root mean square value for $R_n$ which is $R_n \approx 2 \times
10^{-2}$~a.u. for the energies above 7~eV~\cite{Porsev2}.

An interesting effect may take place for neutral thorium if the
frequency of the nuclear transition is smaller than the ionization
potential of the atom.
 The
density of states of Th~I is huge and the resonance situation is
highly probable. This can be used to excite the nucleus in a two-step
process in which a resonance electron state is excited first, then it
gives its energy to excite the nucleus.

\section*{Acknowledgments}

The work was funded in part by the Australian Research Council.

\end{document}